\documentclass[prb,aps,twocolumn,amsmath,amssymb]{revtex4}
\usepackage{graphicx}
\usepackage{dcolumn}
\usepackage{amsmath}
\usepackage{amsbsy}
\usepackage{amsfonts}
\usepackage{bbm}
\usepackage[a4paper]{hyperref}

\begin{document}

\preprint{APS/123-QED}
\title{Fano effect in a ring-dot system with tunable coupling.}

\author{A. Fuhrer,$^{1,2}$ P. Brusheim,$^{1}$ T. Ihn,$^2$ M. Sigrist,$^{2}$ K. Ensslin,$^2$ W. Wegscheider,$^3$ and M. Bichler$^4$}

\affiliation{
$^1$Solid State Physics/Nanometer Consortium, Lund University, Sweden\\
$^2$Solid State Physics Laboratory, ETH Z\"urich, 8093 Z\"urich, Switzerland\\
$^3$Institut f\"ur experimentelle und angewandte Physik, Universit\"at Regensburg, Germany\\
$^4$Walter Schottky Institut, Technische Universit\"at M\"unchen, Germany}

\date{\today}

\begin{abstract}
Transport measurements are presented on a quantum ring that is tunnel-coupled to a quantum dot. When the dot is in the Coulomb blockade regime, but strongly coupled to the open ring, Fano line shapes are observed in the current through the ring, when the electron number in the dot changes by one. The symmetry of the Fano resonances is found to depend on the magnetic flux penetrating the area of the ring and on the strength of the ring--dot coupling. At temperatures above T=0.65\;K the Fano effect disappears while the Aharonov--Bohm interference in the ring persists up to T=4.2\;K. Good agreement is found between these experimental observations and a single-channel scattering matrix model including decoherence in the dot. 
\end{abstract}



\maketitle

\section{Introduction}
Quantum rings and quantum dots are prototype systems for the observation of mesoscopic interference on the one hand and for spectroscopic investigations of discrete level spectra of interacting systems on the other hand.
Ring shaped structures give rise to Aharonov--Bohm (AB) interference \cite{59Aharonov} which can be tuned by applying a magnetic flux through the area enclosed by the ring. Open ring geometries have been used as interferometers, e.g., to study the transmission phase of quantum dots in the Coulomb blockade (CB) regime.\cite{97schuster,01hackenbroich} The discrete level spectrum of Coulomb blockaded quantum dots has been extensively studied using tunneling spectroscopy to probe interaction and spin effects when a gate voltage is used to successively add electrons to such artificial atoms.\cite{97Rkouwenhoven} 

\begin{figure}[tb]
\begin{center}
\includegraphics[width=8cm]{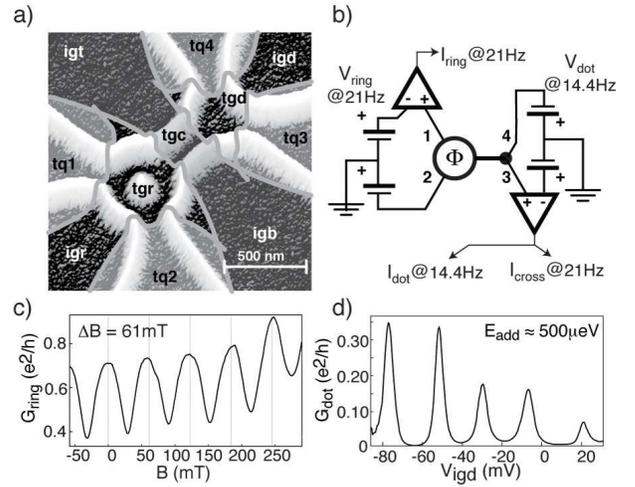}
\caption{(a) Scanning force microscope image of the quantum ring coupled to a quantum dot. The oxide lines (white) impose barriers between the conducting regions of the 2DEG (black), 34nm below the surface. In-plane gate electrodes \emph{igd,igr,igb,igt} as well as a structured top gate \emph{tgd,tgr,tgc,tq1,tq2,tq3,tq4} are used to tune the electron density in the dot and the ring and the coupling between them and to the contacts. (b) Measurement setup used to measure $I_\mathrm{dot}$ and $I_\mathrm{ring}$ simultaneously and under symmetric bias conditions. $I_\mathrm{cross}$ reflects the transmission from the ring to the dot.
(c) Aharonov--Bohm oscillations in the two terminal ring conductance
  $G_\mathrm{ring}$. (d) Coulomb-blockade oscillations in the two-terminal dot
  conductance $G_\mathrm{dot}$  as a function of the in-plane gate voltage $V_\mathrm{igd}$.
}\label{fig1}\end{center}\end{figure}

Interference of a resonant state, e.g. a discrete level of a quantum dot, and a continuum of states, e.g., in a quantum wire or an open ring, typically gives rise to asymmetric line shapes characteristic of the Fano effect.\cite{61fano} A theoretical treatment of the Fano effect was originally developed for scattering cross-sections of nuclei \cite{35Fano,47Feshbach,49adair} and in optical spectroscopy\cite{73cerdeira}, but recently Fano resonnances were also addressed in a multitude of mesoscopic systems \cite{90McEuen,90Bagwell,93ATekman,94nockel,00goeres,02kobayashi,04Akobayashi,04AJohnson}. In mesoscopic transport the energy dependent conductance of a Fano resonance can be written as
\begin{eqnarray}
G(\epsilon)=G_0+\frac{G_1}{1+q^2} \frac{(q+\epsilon)^2}{\epsilon^2+1}.
\label{eqn1}
\end{eqnarray}
Here $\epsilon=(E-E_0)/(\Gamma/2)$ is a dimensionless energy parameter with $E$ the energy of an electron, $E_0$ the energy of the resonance and $\Gamma$ the width of the resonance. The quantity $q$ is known as the \emph{Fano parameter} and determines the shape of the resonance line.\cite{61fano} For $q = \infty$ resonant transmission dominates and the Fano resonance becomes equivalent to a symmetric Breit--Wigner resonance. For $q = 0$ a Breit--Wigner shaped anti-resonance is observed and for $q = \pm1$ the asymmetry is strongest. Theoretically the Fano effect in mesoscopic systems has been discussed for both resonant and open cavities,\cite{98xu,92porod} and in conjunction with rings.\cite{03ueda,96cho,05voo} It has been proposed that Fano resonances in quantum dots could be used as a measure of phase coherence\cite{01clerk} or a means to produce spin polarized currents.\cite{03song} 
Experimentally Fano resonances were observed in the tunneling current through a single cobalt atom on a gold surface.\cite{98madhavan} In Coulomb blockaded quantum dots it was found that the interference of a broad resonance (quasi-continuum) with a narrow resonance can lead to Fano peak shapes.\cite{00goeres,01Zacharia,04schonenberger} A recent experiment investigated a Coulomb blockaded quantum dot side-coupled to a quantum wire and discussed the implications of Coulomb interactions between the dot and the wire in the Fano regime\cite{04AJohnson}. 
Kobayashi et al. further studied the Fano effect in an AB-ring with a quantum dot embedded in one arm of the ring.\cite{02kobayashi} In these experiments the magnetic field allowed them to tune the relative phase between the non-resonant (ring) and the resonant (dot) path, periodically changing the asymmetry of the Fano line shape. Their interpretation required the introduction of a complex $q$-parameter to account for the AB-phase. Similar results were also found in crossed carbon nanotubes with a resonant state at the point where the tubes are touching.\cite{03AKim} 

In these ring systems the Fano effect arises due to the interference of a Breit-Wigner type resonance in one arm of the ring (containing a strongly coupled quantum dot) with a continuum channel in the other arm of the ring.
Here we present transport studies on a structure where a quantum dot in CB-regime is side-coupled to a ring [see Fig.\;\ref{fig1}(a)]. Our structure has a tunable channel between the ring and the dot which permits us to couple coherently the two subsystems while keeping them spatially separated and therefore minimize capacitive effects of the quantum dot on the interference in the ring, as investigated in Ref\;\onlinecite{04AJohnson} and Ref\;\onlinecite{04meier}. In contrast to previous ring systems, our experiment constitutes a tunable Fano scatterer consisting of one arm of the ring and the side coupled dot, which is made to interfere with alternative paths along the second arm of the ring. This allows us to study the interplay between continuous AB-interference in the ring and interference involving a resonant level in the dot. 

The paper is organized as follows: in Section II, the experimental realization of the coupled ring-dot structure is discussed and low temperature transport measurements are presented. In Section III we give a model for the ring-dot system within the scattering matrix formalism and link it to the Fano formula in Eq.\;\ref{eqn1}.  In Section IV model and experimental results are compared and we follow Ref.\;\onlinecite{02benjamin} to model decoherence in the dot due to finite temperatures and coupling to the leads. In Section V we discuss limitations of the model. 

\section{Experimental results}
\subsection{Sample and measurement setup}
\begin{figure}[t]
\begin{center}
\includegraphics[width=3.4in]{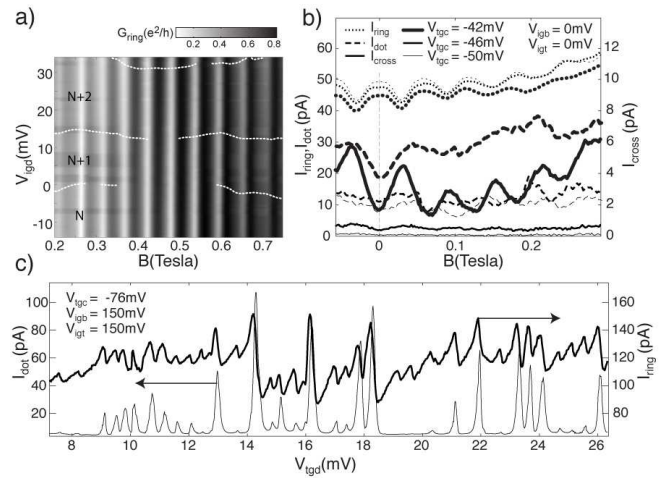}
\caption{(a)~AB oscillations in $G_\mathrm{ring}$ as a function of $V_\mathrm{igd}$ when the channel between the dot and the ring is pinched off. The dashed white lines show the CB-peak positions as determined from the dot conductance $G_\mathrm{dot}$. This indicates that electrostatic coupling is negligibly small.  (b)~Three simultaneously measured currents $I_\mathrm{ring}$ (dotted lines), $I_\mathrm{dot}$ (dashed lines), $I_\mathrm{cross}$ (solid lines), when $V_\mathrm{tgc}$= -50\;mV,-46\;mV,-42\;mV is raised to increase the coupling between the ring and the dot. The thickest line indicates strongest coupling. (c) $I_\mathrm{ring}$ and $I_\mathrm{dot}$ when $V_{tgd}$ is used to tune the electron number of the dot. While $I_\mathrm{dot}$ shows the expected Coulomb oscillations the current through the ring exhibits the typical asymmetric line shapes characteristic of the Fano effect.
}\label{fig2}\end{center}\end{figure}
The coupled ring--dot structure was realized in a Ga[Al]As heterostructure with a high-quality two-dimensional electron gas (2DEG) 34\,nm below the surface. The electron density in the 2DEG was $n_\mathrm{s}=4.8\times 10^{15}$\,m$^{-2}$ and the mobility was $\mu=50$\,m$^2$/Vs at 1.7\,K. Fig.\,\ref{fig1}(a) shows an AFM image of the device defined using AFM-lithography. The 2DEG below the oxide lines is depleted (white lines) and the 2DEG regions at the side of the ring--dot structure are used as in-plane gates (\emph{igt,igr,igb,igd}). In a subsequent step an 8\;nm thin titanium film was evaporated on top of the structure and the AFM was again used to cut this top gate into individual split gates (\emph{tq4,tgd,tq3,tgc,tq1,tgr,tq2}) as indicated by the gray lines. These top gates allow for a unique tunability of the electron number in and the coupling between the two subsystems as well as to the leads. More specifically a narrow top gate stripe (\emph{tgc}) can be used to completely isolate the ring electronically from the dot. Details about the fabrication process may be found in Ref.\;\onlinecite{04sigrist}\,and Ref.\;\onlinecite{02fuhrer}. 

The measurements were performed in a dilution refrigerator with an electron temperature $T=100\;$mK. The measurement setup is shown schematically in Fig.\,\ref{fig1}(b). A symmetric AC bias voltage $\pm V_\mathrm{ring}=5\mu$V was applied to the two leads of the ring at a frequency of 21\;Hz and in the same way $\pm V_\mathrm{dot}=5\mu$V was applied to the two leads of the dot with a frequency of 14.4\;Hz. Using lock-in techniques this allows us to simultaneously detect the following three currents related to the transmission matrix of the four terminal structure (assuming spin degeneracy):
\begin{eqnarray*}
\frac{I_\mathrm{ring}}{V_\mathrm{ring}}&=&\frac{2 e^2 }{h}\left[T_{12}+\frac{1}{2} \left(T_{13}+T_{14}\right)\right]\\
\frac{I_\mathrm{dot}}{V_\mathrm{dot}}&=&\frac{2 e^2 }{h}\left[T_{34}+\frac{1}{2} (T_{31}+T_{32})\right]\\
\frac{I_\mathrm{cross}}{V_\mathrm{ring}}&=&\frac{2 e^2 }{h}(T_{31}-T_{32}).
\end{eqnarray*}

Here T$_{ji}$ is the transmission from lead $i$ to $j$ following the numbering of the gates {\em tqi} in Fig.~\ref{fig1}(a). For negligible coupling between the structures the first two values become equivalent to the two terminal conductance $G_\mathrm{ring}$ of the ring and  $G_\mathrm{dot}$ of the dot respectively since the transmission through the connecting channel
vanishes. A measurement of $G_\mathrm{ring}$ and $G_\mathrm{dot}$ when the central gate pinches off the channel between the structures is shown in Fig.\;\ref{fig1}(c) and (d). The ring conductance exhibits pronounced AB-oscillations as a function of a magnetic field $B$ applied perpendicular to the sample and the period $\Delta B=61\;$mT is in agreement with adding single flux quanta to the area enclosed by the ring with an average radius $r_0=$135\;nm. The AB-oscillations persist up to $T\approx4.2\;$K. This is slightly lower than in previous measurements of similar two-terminal ring structures where a phase coherence length $\ell_\phi$ of more than a micron was found at 4.2\;K.\cite{03Cihn}
The conductance $G_\mathrm{dot}$ shows CB-resonances as a function of $V_\mathrm{igd}$ applied to the in-plane gate electrode \emph{igd} closest to the dot. From measurements of Coulomb diamonds we find an average addition energy $E_\mathrm{add}\approx500\;\mu$eV for the dot. 
Our set-up allows us to simultaneously detect the resonance position in the dot and the interference pattern in the ring current, without changing the coupling or any additional gate voltages. Furthermore, the magnetic field is used to tune the interference between the phase sensitive ring and the dot which, when side coupled to one arm of the ring, will act as a Fano scatterer (see below). 

\subsection{Coupling the ring and the dot}
Before coupling the ring and the dot coherently we assess the magnitude of the crosstalk between the structures due to the cross-coupling of the gate electrodes and direct Coulomb interaction. To this end $V_\mathrm{tgc}$ was set to -75\;mV in order to deplete the channel between the two structures. Figure\;\ref{fig2}(a) shows a measurement of AB-oscillations in $G_\mathrm{ring}$ when $V_\mathrm{igd}$ is used to tune the electron number $N$ in the dot. The white dotted lines show the CB-peak positions in the dot. The unperturbed AB-oscillations in Fig.\;\ref{fig2} indicate that unlike recently reported results \cite{04AJohnson,04meier} we are in a regime where the direct Coulomb interaction between the two structures is negligibly small. Furthermore, the gate on the dot changes neither the characteristic amplitude nor the phase of AB-oscillations in the ring over a range $\Delta V_{\mathrm{igd}}\geq45\;$mV. We find that the same is true when changing $V_\mathrm{tgd}$ applied to the top gate of the dot. The top gate electrodes \emph{tgr} and \emph{tgd} respectively allow us to tune the electron density in the ring or electron number in the dot over a large range without changing the coupling to source and drain strongly. However, where possible we restrict ourselves to tuning the in-plane gates, since in this case we find the structure to be more stable and the gate voltages lead to fewer charge rearrangements.

In order to couple the two structures we increase $V_\mathrm{tgc}$. Figure\;\ref{fig2}(b) shows the three currents $I_\mathrm{ring}$(dotted line), $I_\mathrm{dot}$(dashed line) and $I_\mathrm{cross}$(solid line) where thicker lines correspond to less negative $V_\mathrm{tgc}$. With stronger coupling, $I_\mathrm{cross}$ increases from negligible current at $V_\mathrm{tgc}=-50\;$mV to a value around 4\;pA with clear AB-oscillations at $V_\mathrm{tgc}=-42\;$mV.
In this regime $I_\mathrm{ring}$ exhibits larger deviations from the perfect symmetry as a function of magnetic field $B$, an effect that is expected for the transition from a two-terminal to multi-terminal system.\cite{86Bbuttiker} The current $I_\mathrm{dot}$ is not significantly modified by the increase in $V_\mathrm{tgc}$ except that the source-drain coupling of the dot is influenced slightly which leads to a larger current for the strongest coupling. 
\begin{figure}[t]
\begin{center}
\includegraphics[width=3.4in]{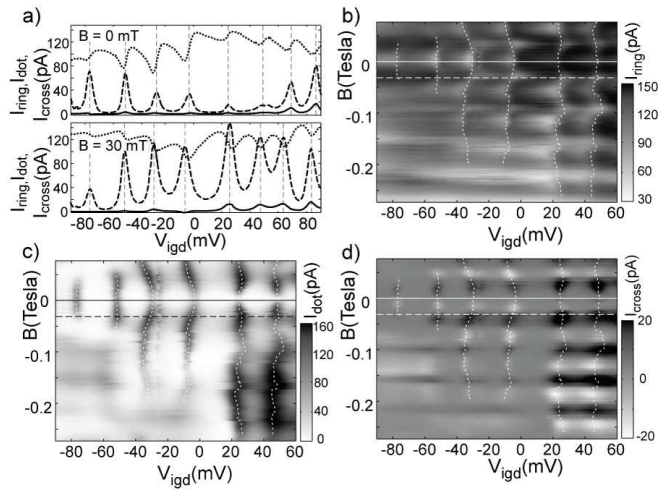}
\caption{(a) Line traces as a function of $V_\mathrm{igd}$ of the currents $I_\mathrm{ring}$ (dotted lines), $I_\mathrm{dot}$(dashed lines) and $I_\mathrm{cross}$ (solid lines) at $B=0\;$mT (top panel) and $B=$30 mT (bottom panel). The vertical dashed lines mark the peak positions.  Curves are slightly offset at $V_\mathrm{igd}$=-25\;mV to correct for a parametric charge rearrangement at that gate value. (b) - (d) Gray scale plots of the currents $I_\mathrm{ring}$ (a), $I_\mathrm{dot}$(b) and $I_\mathrm{cross}$(c) as a function of $B$ and $V_\mathrm{igd}$. The white vertical lines are fits to the CB-peak positions in (c). }\label{fig3}\end{center}\end{figure} 

Starting from this strongly coupled situation, we have optimized the amplitude of the AB-oscillations in the ring and their coupling to the CB-resonances in the dot by changing several gate voltages. Note that specifically $V_\mathrm{igt}$ and $V_\mathrm{igb}$ were increased from 0\;mV to 200\;mV in order to optimize the interconnecting channel from the ring to the dot. In the following, more negative $V_\mathrm{tgc}$ values are therefore required to close this channel. Figure\;\ref{fig2}(c) shows a simultaneous measurement of $I_\mathrm{ring}$ and $I_\mathrm{dot}$ when $V_\mathrm{tgd}$ is tuned over a large range. While $I_\mathrm{dot}$ shows clear CB-oscillations the ring current depends strongly on the level alignment in the dot. When the dot is tuned through a resonance peak  with increasing $V_\mathrm{tgd}$, $I_\mathrm{ring}$ is reduced by up to a factor of two. 
This is a manifestation of the Fano effect. It has been shown previously that a dot which is side-coupled to a wire is a typical Fano system.\;\cite{04Akobayashi}
Here we probe the properties of such a Fano scatterer by embedding it in a two terminal ring structure.

Figure\;\ref{fig3}(a) shows curve traces of  $I_\mathrm{ring}$ and $I_\mathrm{dot}$ as a function of $V_\mathrm{igd}$ for $B=0$\;mT and after half an AB-period ($B=30\;$mT). The vertical lines mark the positions of the CB-resonances as determined from peak fits to $I_\mathrm{dot}$ with a Lorentzian line shape. Increasing the flux through the ring by half a flux quantum changes $I_\mathrm{ring}$ from a behavior where a dip occurs at each gate voltage before the dot resonance to a situation where a similar asymmetric dip occurs after the dot resonance.
Figures\;\ref{fig3}(b)-(d) show $I_\mathrm{ring}$, $I_\mathrm{dot}$ and $I_\mathrm{cross}$ in gray-scale plots as a function of $V_\mathrm{igd}$ and $B$. The dips in $I_\mathrm{ring}$ [bright areas in Fig.\;\ref{fig3}(b)] continue to oscillate back and forth for each additional flux quantum through the ring until at $B\ge150\;$mT normal AB-oscillations are recovered. The vertical white lines are again the peak positions determined from fits to $I_\mathrm{dot}$ shown in Fig.\;\ref{fig3}(c). The slight peak shifts as a function of magnetic field are due to the influence of the magnetic field on individual orbital levels of the dot. From the evolution of the peak amplitude and position, we find that neighboring peaks show highly correlated behavior over large magnetic field ranges indicating spin pairing in the dot for the six peaks that are shown. As expected $I_\mathrm{cross}$ [Fig.\;\ref{fig3}(d)] is different from zero only when $I_\mathrm{dot}$ is finite. $I_\mathrm{cross}$ which flows through both structures oscillates around zero with the AB-period as a function of $B$. This is in contrast to $I_\mathrm{ring}$ which has a significant $B$-independent background. We explain this by the fact that $I_\mathrm{cross}$ measures the difference between the two transmissions $T_{31}$ and $T_{32}$ and the $B-independent$-contribution therefore cancels out if the structure geometry and the couplings are ideally symmetric. Due to the three terminals the different AB-geometry for $T_{31}$ and $T_{32}$ retains a net AB-effect.


\section{Theoretical Model}
\subsection{A Quantum Dot Resonator}
\begin{figure}[t]
\begin{center}
\includegraphics[width=3.4in]{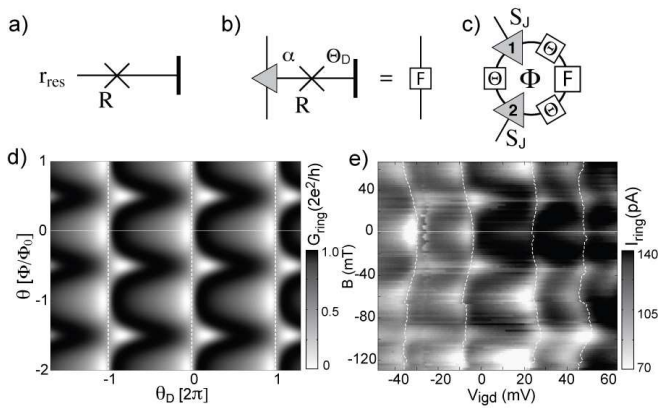}
\caption{(a) Stub resonator with barrier and reflector as a model for a quantum dot. (b) Fano scatterer realized with a dot side-coupled to a wire. (c) Ring-dot structure composed of a Fano scatterer embedded in a two terminal AB-ring. (d) and (e) Comparison of the scattering model with $I_\mathrm{ring}$. (d) Model calculation for -2 to +1 flux quanta passing through the ring and over a
  change of $6\pi$ in the dot phase $\theta_\mathrm{D}$. The
  parameters used were $q=1$, $\theta_\mathrm{R}=0.61$, $R=1/2$ and all leads
  to the ring are assumed to be ideal beam splitters. (e)
  Section of $I_\mathrm{ring}$ for two spin paired states. The white dashed
  lines are the peak positions extracted from fits to $I_\mathrm{dot}$.
  }\label{fig4}\end{center}\end{figure}

We model the coherent quantum dot in the single particle, single level approximation as a closed-end resonator consisting of a 1D
wire with a barrier and a perfect reflector as shown in Fig.~\ref{fig4}(a). The unitary scattering matrix of a
general barrier is given by 
\begin{eqnarray}
  S_R &=& \left(%
  \begin{array}{cc}
    r & t'\\
    t & r'\\
  \end{array}%
  \right) \nonumber \\
  &=& \left(%
  \begin{array}{cc}
    e^{i\alpha}\sqrt{R} & e^{i(\alpha+\kappa)/2}\sqrt{1-R}\\
    e^{i(\alpha+\kappa)/2}\sqrt{1-R} & -e^{i\kappa}\sqrt{R}\\
  \end{array}%
  \right),
\end{eqnarray}
where $R$ is the reflection probability and $\alpha,\kappa$ are free phase parameters. The reflection amplitude of such a
dot is found from a sum over all Feynman paths
\begin{equation}
 r_\mathrm{res} = r + t\lambda r_\mathrm{t} \lambda\left[ \sum_{n=0}^\infty
 (r'\lambda r_\mathrm{t}\lambda)^n \right]t = r+\frac{t\lambda
 r_\mathrm{t}\lambda t}{1-r'\lambda r_\mathrm{t}\lambda}.
\end{equation}
Here $\lambda = e^{i\beta}$ describes the propagation between the barrier and
the reflector and $r_\mathrm{t}=-e^{i\delta}$ is the reflection amplitude at
the reflector. By introducing the phase
$\theta_\mathrm{D}=\kappa+2\beta+\delta$ which an electron accumulates during a
round trip in the dot, the total reflection
amplitude becomes
\begin{equation}
  r_\mathrm{res} = -e^{i\alpha}\frac{e^{i\theta_\mathrm{D}}-\sqrt{R}}{1-\sqrt{R}e^{i\theta_\mathrm{D}}}.
\end{equation}  
This can be rewritten as
\begin{equation}
  r_\mathrm{res} = -e^{i\alpha}\frac{1+\frac{i}{\gamma}\tan\frac{\theta_\mathrm{D}}{2}}{1-\frac{i}{\gamma}\tan\frac{\theta_\mathrm{D}}{2}}=-e^{i\alpha}\frac{1+i\epsilon}{1-i\epsilon}=-e^{i(\alpha+2\arctan \epsilon)}
\end{equation}
with $\gamma=(1-\sqrt{R})/(1+\sqrt{R})$ and
$\epsilon=\frac{1}{\gamma}\tan\frac{\theta_\mathrm{D}}{2}$. For this closed-end structure, the amplitude is restricted to the
unit circle and resonances arise when $\theta_\mathrm{D}=2\pi n$, $n\in\mathbbm{Z}$.

The dot resonator is then side-coupled to a wire through a three terminal junction as
seen in Fig.~\ref{fig4}(b). The transmission and
reflection amplitudes of the wire are
\begin{eqnarray}
  t_\mathrm{W} &=& t_\mathrm{d} + b_1 r_\mathrm{res} b_2, \nonumber \\
  r_\mathrm{W} &=&  r_\mathrm{d} + b_1 r_\mathrm{res} b'_1,\\
\end{eqnarray}
where the amplitudes $t_\mathrm{d}$ and $r_\mathrm{d}$ describe the direct transmission from the
entrance to the exit of the wire and the direct reflection at the entrance of the
wire respectively. $b_1$ and $b'_1$ are the amplitudes for transmission from
the entrance into the dot and from the dot to the entrance respectively, and $b_2$ is the amplitude for transmission from the
dot into the exit of the wire. The side-coupled dot structure can hence be mapped onto a $2\times2$ transfer matrix
\begin{equation}
  F =
  \frac{1}{t_\mathrm{W}}\left(\begin{array}{cc}t_\mathrm{W}^2-r_\mathrm{W}^2 &
  r_\mathrm{W}\\ -r_\mathrm{W} & 1 \end{array}\right).\label{FT}
\end{equation}
as indicated in Fig.~\ref{fig4}(b). 
The transmission probability of the wire with the side-coupled dot can then be shown to have the Fano form
\begin{equation}
  T_\mathrm{W} = |t_\mathrm{W}|^2 = 
  (t_\mathrm{d}-b_1b_2)^2+\frac{4t_\mathrm{d}b_1b_2}{1+q^2}\frac{(q+\epsilon)^2}{1+\epsilon^2},
  \label{TW}
\end{equation}
where $q=\tan(\alpha/2)$ is the Fano parameter. For $b_1, b_2\neq 0$ $T_\mathrm{W}$ produces Fano
line shapes. This becomes clear when $\epsilon$ is approximated close to a resonance at energy $E_p$, neglecting any energy dependence of $\gamma$. It follows that $\epsilon=\frac{1}{\gamma}\tan\frac{\theta_D}{2}\approx\frac{E-E_p}{\Gamma}$ with $\Gamma=2\gamma\left(\frac{d\theta_D}{dE}\right)^{-1}$ giving the link to the Fano formula in Eq.\;\ref{eqn1}. Note, that the Fano parameter $q$ in Eq.\;\ref{TW}, and through it the line shape, is uniquely determined by the direct reflection phase $\alpha$ of the barrier in the side-coupled dot structure, i.e., $\alpha=\pm\pi$ produces a symmetric resonance peak
($q=\pm\infty$), $\alpha=0$ produces a symmetric 
dip in the transmission ($q= 0$), and $\alpha=\pm\pi/2$ produce maximally
asymmetric line shapes ($q= \pm1$). 

In the following we assume the three terminal junction to be a symmetric beam splitter which is described by the
unitary scattering matrix~\cite{84buttiker} 
\begin{eqnarray}
  S_J=\left(%
  \begin{array}{ccc}
    0 & \frac{1}{\sqrt2} & \frac{1}{\sqrt2}\\
    \frac{1}{\sqrt2} & -\frac{1}{2} & \frac{1}{2}\\
    \frac{1}{\sqrt2} & \frac{1}{2} & -\frac{1}{2}\\
  \end{array}%
  \right).
\end{eqnarray}
Using this for the coupling between the dot and the wire we get $t_\mathrm{d}=b_1b_2=b_1b'_1=1/2$ resulting in a vanishing
constant term in the transmission probability in Eq.~\ref{TW}. It follows that
\begin{eqnarray}
  t_\mathrm{W} & = &
  \frac{1}{2}(1+r_\mathrm{res})=\frac{1}{i+q}\frac{q+\epsilon}{1-i\epsilon},
  \nonumber \\
  r_\mathrm{W} & = & -\frac{1}{2}(1-r_\mathrm{res}) = \frac{i}{i+q}\frac{q\epsilon-1}{1-i\epsilon}.\nonumber
\end{eqnarray}
A \emph{Fano scatterer} is therefore characterized by the transfer matrix $F$ in Eq.\;\ref{FT}, with the two parameters $q$ and $\epsilon$.
 \begin{figure*}[th!]
 \begin{center}
\includegraphics[width=7.1in]{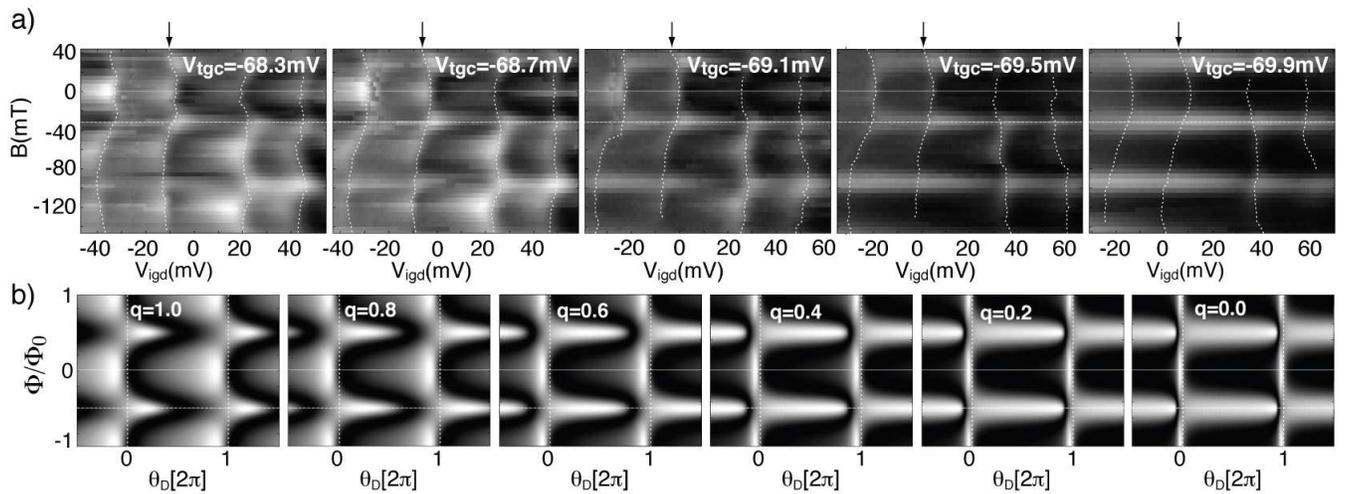}
\caption{Comparison of the scattering model with $I_\mathrm{ring}$ for different coupling strengths. (a) 
  $I_\mathrm{ring}$ as a function of $V_\mathrm{igd}$ and $B$ for two spin paired states and decreasing values of $V_\mathrm{tgc}$. The white dashed
  lines are the peak positions extracted from fits to $I_\mathrm{dot}$. As $V_\mathrm{tgc}$ is decreased from left to right the gate axis is shifted to compensate for the $V_\mathrm{tgc}$-induced shift in peak position. (b) $G_\mathrm{21}$ for -1 to +1 flux quanta passing through the ring and over a
  range of $4\pi$ in $\theta_\mathrm{D}$. The parameters used are $\theta_\mathrm{R}=0.61$, $q=1.0, 0.8, 0.6, 0.4, 0.2, 0$ and $R=1/2$. $G_\mathrm{21}$ ranges from 0 (white) to $2e^2/h$ (black)}\label{fig5}\end{center}
\end{figure*}

\subsection{Ring-Dot Structure} 
The ring-dot structure in the experiment can now be
modeled by placing the transfer matrix $F(\epsilon,q)$ in one of the arms of the ring as
shown in Fig.~\ref{fig4}(c). The ring is attached to two ideal semi-infinite
leads at the junctions 1 and 2. These junctions together with the transfer
matrix $F$ partitions the ring into three sections of equal length. When an electron wave traverses each section it acquires the Fermi phase
 $\theta_\mathrm{R}=k_F 2\pi r_0/3$, where $k_F$ is the Fermi wave-number, as well as a magnetic phase $\theta=\pm\Phi/(3\Phi_0)$ (opposite signs for clockwise and anticlockwise propagation) where
 $\Phi$ is the magnetic flux enclosed by the ring and $\Phi_0=h/e$ is the magnetic
 flux quantum. These phases enter the transfer matrix for each ring section 
\begin{eqnarray}
  \Theta=\left(%
  \begin{array}{cc}
    e^{-i(\theta-\theta_\mathrm{R})} & 0\\
    0 & e^{-i(\theta+\theta_\mathrm{R})}\\
  \end{array}%
  \right)
\end{eqnarray} 
as indicated in Fig.\;\ref{fig4}(c). 
With the boundary conditions on the incoming amplitudes in the source and
drain leads $a_1=1$ and $a_2=0$, one finds the two-terminal conductance from the Landauer formula
\begin{equation}
  G_\mathrm{21} = \frac{2e^2}{h}T_\mathrm{21} = \frac{4e^2}{h}\left|c_\mathrm{21}+c'_\mathrm{21}\right|^2,
\end{equation}
with
\begin{eqnarray}
  &\left(
  \begin{array}{c}
    c_\mathrm{21}   \\
    c'_\mathrm{21}  \\
  \end{array}
  \right)& = \Pi^{-1} \Theta F \Theta \tau, \nonumber \\
  &\Pi& = \mathbbm{1} - \Theta F \Theta t_1 \Theta t_2, \nonumber \\
  &t_1& = t_2 = \left(
  \begin{array}{cc}
    0 & -1 \\
    1 & 2 \\
  \end{array}\right),
\end{eqnarray}
and
\begin{equation}
  \tau = {\sqrt2}\left(
  \begin{array}{c}
     1 \\
    -1 \\
  \end{array}
  \right).
\end{equation}

\section{Comparison with the experiment}
\subsection{Tuning q with a gate voltage}
For comparison with the experimental data we set $\theta_\mathrm{R} = 0.61$, a value which gives pronounced
AB-oscillations in the ring while suppressing higher harmonics when the dot is
decoupled (R=1). Figure\;\ref{fig4}(d) shows the calculated conductance $G_\mathrm{21}$ as a function of $\theta_D$ and the number of flux quanta penetrating the ring with $\alpha=\pi/2\;(q=1)$ and $R=1/2$. For comparison, a similar section of the measured $I_\mathrm{ring}$  as a function of B and $V_{igd}$ is shown in Fig.\;\ref{fig4}(e). 
Very good agreement is found, indicating that our simple model quite accurately describes the experimental situation of the coupled ring-dot structure.  This agreement is robust against variations in $R$ but sensitive to changes in $\alpha$, since $\alpha$ is linked to the Fano factor $q$. We determine $q=1\pm0.1$ in agreement with the experimental procedure where we optimized the coupling between the ring and the dot to result in most asymmetric Fano line shapes. 
The sign of the asymmetric line shape in $I_\mathrm{ring}$ is periodically tuned by the flux penetrating the ring. This behavior is qualitatively similar to recently reported results in Ref.\;\onlinecite{02kobayashi} where a complex q was introduced in order to accommodate for the magnetic field dependence of the Fano line shapes. However, we have not been able to transform Eq.\;\ref{TW} into a form where the magnetic flux through the ring $only$ enters as a phase $\Phi/\Phi_0$ in a complex $q$-parameter and we therefore restrict ourselves to a real $q$-parameter reflecting only the Fano effect of one arm of the ring with the side coupled dot. The interference properties of this Fano scatterer are then probed by AB-interference in the ring. This is different to Ref.\;\onlinecite{02kobayashi} where the whole ring structure with a dot embedded in one arm is considered as a Fano scatterer. 

 The $q$-parameter is indicative of the coupling between the dot and the ring and to further strengthen the agreement between model and experiment we tune $V_{tgc}$ successively more negative to reduce the coupling and thereby tune the $q$-parameter. Figure\;\ref{fig5}(a) shows $I_\mathrm{ring}$ as a function of $B$ and $V_{igd}$ for decreasing values of $V_{tgc}$ from left to right. 
As the coupling is decreased $I_\mathrm{ring}$ becomes more strongly suppressed at $B=-30$\;mT while the dips (white areas) at $B=0$\;mT seem to disappear. Within our model, this can be understood assuming that a decrease in $V_{tgc}$ reduces the phase pick-up $\alpha$ of the electrons traversing the interconnecting channel and therefore decreases $q$ from its initial value of 1. In addition, the reflection probability R is likely to increase. Figure\;\ref{fig5}(b)
shows $G_\mathrm{21}$ for the same parameters as in Fig.\;\ref{fig4}(a) while varying $q$ from 1 to 0. Again we find good qualitative agreement with the experiment except for very small $q$-values where it is difficult to determine which q matches the experimental situation best. The last plot on the right shows $G_\mathrm{21}$ for $q=$0 where AB-like oscillations are found except very close to the resonance of the dot. 
This transition is more complete in the experimental data, where for $V_\mathrm{tgc}=-69.9$\;mV only small dips in $I_\mathrm{ring}$ are found.

In Fig.\;\ref{fig6}(a) traces of $I_\mathrm{ring}$ and  $I_\mathrm{dot}$ as a function of $V_\mathrm{igd}$ at $B=0$\;mT (solid lines) and $B=30\;$mT (dashed lines) are plotted for $V_\mathrm{tgc}=$-68.3,-68.7,-69.1,-69.5, and -69.9\;mV. The traces have been taken across the Coulomb resonance marked with arrows in Fig\;\ref{fig5}(a) and for each value of $V_\mathrm{tgc}$ they were shifted in $V_\mathrm{igd}$ such that the peak maximum in $I_\mathrm{dot}$ (only $B=0$\;mT shown) comes to lie at $\Delta V_\mathrm{igd}=0\;$mV. The topmost solid (lowermost dashed) line respectively mark an AB-maximum(minimum) for $V_\mathrm{tgc}=$-69.9\;mV which is the most weakly coupled situation and corresponds to a $q<0.4$. Here the dip(peak)in $I_\mathrm{ring}$ close to the resonance position are more clearly visible. The behavior in Fig.\;\ref{fig5}(b) is qualitatively different in the calculated $G_\mathrm{21}$ where even for $q=0$ the dip close to resonance at $B=0$ yields a sharp transmission zero for the parameters that were chosen. We will argue in the following that this discrepancy between model and experiment can be understood when taking into account decoherence in the dot. In contrast to the model, the experimental quantum dot is connected to two contacts which will lead to partial phase randomization of the electron paths visiting the quantum dot.\cite{84buttiker}

\subsection{Decoherence}
\begin{figure}[t]
\begin{center}
\includegraphics[width=3.4in]{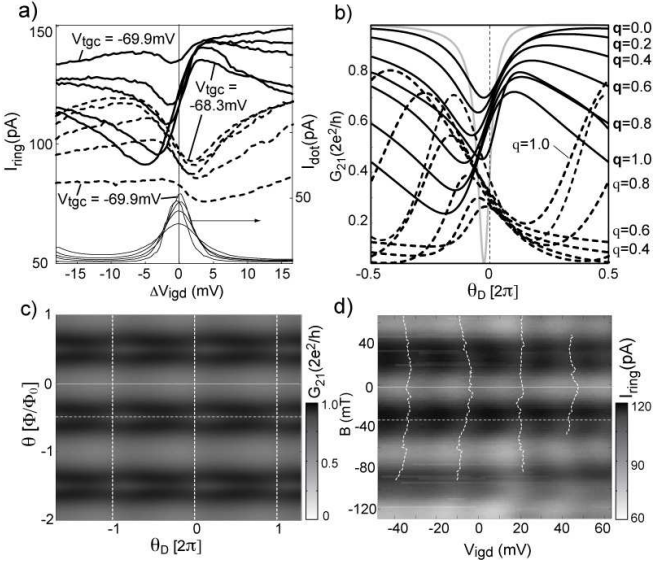}
\caption{(a) $I_\mathrm{ring}$ at $T$=0.1\;K  for decreasing gate coupling tuned by $V_\mathrm{tgc}=$-68.3,-68.7,-69.1,-69.5, and $-69.9\;$mV. The curves are measured over the peak marked with an arrow in Fig.\;\ref{fig5} and for $B=0\;$mT (solid lines), $B=-30\;$mT (dashed lines). For each value of $V_\mathrm{tgc}$ we subtract an offset in $V_\mathrm{igd}$ in order for the peak in $I_\mathrm{dot}$ to be at $V_\mathrm{igd}=0\;$mV. (b) $\tilde{G}_\mathrm{21}$ for $q=1,0.8,0.6,0.4,0.2$ and 0 including decoherence in the dot with a decoherence rate $\ell/\ell_\phi=1/2$. (c)
  $\tilde{G}_\mathrm{21}$ with strong decoherence in the dot using $\ell/\ell_\phi=2$. (d) $I_\mathrm{ring}$ as a function of $V_\mathrm{igd}$ and $B$ at $T$=0.65\;K.}\label{fig6}\end{center}\end{figure}
Since the observed Fano resonances rely on coherent
transport through the whole structure, decoherence at finite temperatures due to coupling of the structure to the environment, e.g. the source and drain leads of the dot, is important. 
We have measured the temperature dependence of the Fano line shapes in 
$I_\mathrm{ring}$ and found that at $T=0.65$\;K the
AB-oscillations in the ring persist while the Fano behavior is strongly
suppressed. This is in agreement with a picture where the paths that pass
through the dot are most strongly affected by decoherence due to the fact that
such a path is longer than the direct paths in the ring which don't traverse the
dot. Furthermore, the dot paths will be influenced by decoherence induced by the two
contacts attached to it.

We follow
Ref.\;\onlinecite{02benjamin} in order to include decoherence in the dot by the
modification of the phase $\theta_D \rightarrow \theta_D +
i\ell/\ell_{\phi}$ where $\ell$ is the length of the path through the dot and
$\ell_\phi$ is the phase coherence length. This introduces a damping of the coherent channel passing
through the dot. The "absorbed amplitude" is re-injected into the incoherent
channel by introducing an additional term in the two-terminal conductance
\begin{equation}
  \tilde{G}_\mathrm{21} = G_\mathrm{21} + \frac{2e^2}{h}\frac{(1 - R_\mathrm{11} - T_\mathrm{21})(1 - R_\mathrm{22} - T_\mathrm{21})}{(1 - R_\mathrm{11} -
  T_\mathrm{21} + 1 - R_\mathrm{22} - T_\mathrm{12})},
\end{equation}
where $T_{ji}(R_{ji})$ is the transmission(reflection) probability for
injection in lead $i$ and detection in lead $j$ and is given by 
\begin{eqnarray}
  R_\mathrm{11} &=& \left|\sqrt{2}\left(c_\mathrm{11} + c'_\mathrm{11}\right) - 1\right|^2, \nonumber \\
  T_\mathrm{12} &=& 2\left|c_\mathrm{12} + c'_\mathrm{12}\right|^2, \nonumber \\
  R_\mathrm{22} &=& \left|\sqrt{2}\left(c_\mathrm{22} + c'_\mathrm{22}\right)-1\right|^2,
\end{eqnarray}

with

\begin{eqnarray}
  \left(%
  \begin{array}{c} 
    c_\mathrm{11} \\
    c'_\mathrm{11} \\
  \end{array}%
  \right) &=& \Gamma^{-1} \Theta t_2 \Theta F \Theta \tau, \nonumber \\
  \left(%
  \begin{array}{c} 
    c_\mathrm{12} \\
    c'_\mathrm{12} \\
  \end{array}%
  \right) &=& \Gamma^{-1} \Theta \tau, \nonumber \\ 
\left(%
  \begin{array}{c} 
    c_\mathrm{22} \\
    c'_\mathrm{22} \\
  \end{array}%
  \right) &=& \Pi^{-1} \Theta F \Theta t_1 \Theta \tau, 
\end{eqnarray}
and
\begin{equation}
  \Gamma = \mathbbm{1} - \Theta t_2 \Theta F \Theta t_1.
\end{equation}
The re-injection asserts probability conservation in the structure and hence the
symmetries required by the two-terminal Onsager
relation. 

Figure\;\ref{fig6}(b) shows $\tilde{G}_\mathrm{21}$ for $q=1.0$, $0.8$, $0.6$, $0.4$, $0.2$, and $0$ taking into account a finite phase coherence length $\ell_\phi=4\ell$. The solid lines are again for $\Phi/\Phi_0=0$ and the dashed lines for $\Phi/\Phi_0=1/2$. The gray solid line shows the line shape for $q=0$ in the fully coherent case for comparison. Decoherence leads, as expected, to broadening of the Fano line shapes and to strong amplitude reduction, a fact which has been suggested to be used as a measure for decoherence in a quantum dot in the Fano regime\;\cite{01clerk}. 
The close resemblance to the measured data at $T=100$\;mK [Fig.\;\ref{fig6}(a)] suggests that even at these low temperatures the contacts to the quantum dot lead to decoherence of paths entering the quantum dot from the ring, hence, a modification of
the Fano effect which we measure in $I_\mathrm{ring}$.

However, the deviation from the fully coherent limit is strongest in the regime of weak coupling between the two structures close to $q=0$. The experimental traces show a reduction in oscillation-amplitude (separation between solid and dashed curve) of about 20\% close to resonance, while $\tilde{G}_\mathrm{21}$ exhibits a reduction which is more than twice as large. We believe that in the real structure both the reflection probability $R$ and the decoherence rate increase as the coupling between the ring and the dot is decreased. For weaker coupling, the time an electron spends in the dot becomes larger and this means that close to the dot resonance its probability for tunneling into one of the contacts of the dot becomes larger. We have adjusted the parameters in the model accordingly and find good agreement for $R=3/4$ and $\ell=\ell_\phi$ (not shown). Note, that the cross-coupling of $V_\mathrm{tgc}$ has the opposite effect and the Coulomb peak width in $I_\mathrm{dot}$ decreases for decreasing $V_{tgc}$.
 
For $q=1$ the effect of decoherence due to the dot contacts is small since we expect strong coherent coupling between the two structures. This can be seen from the good agreement between $\tilde{G}_\mathrm{21}$ and $I_\mathrm{ring}$ for this situation.

When the temperature is increased decoherence becomes important also for the $q=1$  case and the Fano effect is almost completely lost at $T=0.65\;K$. Figure\;\ref{fig6}(c) shows $\tilde{G}_\mathrm{21}$ for $R=3/4$, $\theta_R=0.61$, $q=1$ and $\ell_\phi=\ell/2$. These parameters were chosen such as to resemble the data in Fig.\;\ref{fig6}(d) measured at $T=0.65\;K$. Using the dimensions of the ring dot structure we estimate a phase coherence length $\sim 300$\;nm at this temperature. The decoherence of paths through the dot strongly suppresses the asymmetric line shapes, leading to an
AB pattern in the conductance. This clearly links the observed asymmetric line shapes to the interaction between discrete states formed in the dot and continuous states of the ring and further strengthens our explanation in terms of the Fano effect.

\section{Discussion}
The scattering model clearly reproduces all the characteristic features of the experimental data but has a few obvious limitations. While we restrict decoherence to the dot, a more accurate model would include both thermal averaging as well as decoherence in the dot, the ring and the interconnecting arm. Using thermal averaging alone we were not able to reproduce the experimental data at finite temperature and the inclusion of decoherence in the ring and the interconnecting arm did not improve the model. We therefore restricted ourselves to decoherence as discussed above and speculate that the additional leads on the dot are the main source for decoherence of the Fano effect.

The model only treats a two-terminal geometry which neglects effects of the two contacts on the dot in terms of the symmetry of the current as a function of $B$. This means that the model also assumes that there are no net currents flowing from the dot to the ring ($T_{\mathrm{1\leftarrow dot}}=0$). From the simultaneous measurements of $I_\mathrm{cross}$ we estimate this contribution to be about two times smaller than the dips (peaks) in $I_\mathrm{ring}$ for $q=0$ and we are therefore led to believe that it has little influence on $I_\mathrm{ring}$. However, where the current through the dot is suppressed at larger magnetic fields [e.g., in the lower left corner in Fig.\;\ref{fig3}(c)] AB-periodic features are found in $I_\mathrm{dot}$ indicating that either the coherent coupling between the two systems also influences $I_\mathrm{dot}$ or that there is a considerable AB-modulated net current flowing from the ring leads into the dot.

In a quantum dot the addition energy between two resonances is not only determined by space quantization effects but is enhanced by Coulomb interactions between the electrons on the dot. This is not included in our calculation and could explain the slightly stronger modulation as a function of gate voltage in the experimental data when comparing with the model at $T=0.65\;$K in Fig.\;\ref{fig4}(c)-(d).
While we think that our model accurately describes the regime where the Fano effect is strong, we find that in the experimental data at $T=0.65\;$K the dips in $I_\mathrm{ring}$ are closer to the actual peak positions of the dot than in the model. We have shown that the influence of cross-coupling from the dot is negligible in the completely decoupled system (see also Ref\;\onlinecite{04meier}). However for $V_\mathrm{tgc}=$-69.9 mV the channel between the two structures is not pinched off and Coulomb interactions could be stronger and possibly lead to modified Fano line shapes close to the resonance condition of the dot.\cite{04AJohnson} 

We find that spin paired peaks involving subsequent filling of the same orbital do not only show similar behavior in  $I_\mathrm{dot}$ but also lead to a similar interference pattern in  $I_\mathrm{ring}$ as evident, e.g., in Fig.\;\ref{fig4}(b). We conclude, that spin is not important for the observed interference effect and Kondo correlations\cite{05ASato} are therefore negligible.

In conclusion, we have demonstrated that the current through a ring with a
coherently side coupled dot leads to asymmetric line shapes in the current through the ring as a function of a gate voltage which tunes the discrete dot levels. This is indicative of the Fano effect and we have shown that the symmetry of the observed Fano line shapes can be tuned by applying a magnetic field. The $q$-parameter was adjusted with a gate voltage tuning the coupling between the ring and the dot. Good agreement between the data and a single-channel scattering matrix model allows us to identify our structure with a Fano scatterer embedded in a two-terminal AB-ring. Comparison with this model also lets us to identify the additional leads on the dot as the dominant source of decoherence for the Fano effect. Raising the temperature to $T=0.65\;$K the Fano effect almost completely disappears while AB-interference in the ring persists to above $T=4.2\;$K.

Financial support from the Swiss Science Foundation (Schweizerischer Nationalfonds) is gratefully acknowledged. We also acknowledge valuable discussions with Hongqi Xu. 

\bibliographystyle{apsrev}

\end{document}